# Availability Aware Continuous Replica Placement Problem

Abdullah
Center for Mobile Cloud Computing Research (C4MCCR), Faculty of Computer Science and IT
University of Malaya,
Kuala Lumpur, Malaysia
abdullahyousafzai@siswa.um.edu.my

Abdullah Gani
Center for Mobile Cloud Computing Research (C4MCCR), Faculty of Computer Science and IT
University of Malaya,
Kuala Lumpur, Malaysia
abdullah@ um.edu.my

Rafidah Md Noor
Faculty of Computer Science and IT
University of Malaya,
Kuala Lumpur, Malaysia
rafidah@ um.edu.my

***Abstract*— Replica placement (RP) intended at producing a set of duplicated data items across the nodes of a distributed system in order to optimize fault tolerance, availability, system performance load balancing. Typically, RP formulations employ dynamic methods to change the replica placement in the system potentially upon user request profile. Continuous Replica Placement Problem (CRPP) is an extension of replica placement problem that takes into consideration the current replication state of the distributed system along with user request profile to define a new replication scheme, subject to optimization criteria and constraints. This paper proposes an alternative technique, named Availability Aware Continuous Replica Placement Problem (AACRPP).AACRPP can be defined as: Given an already defined replica placement scheme, a user request profile, and a node failure profile define a new replication scheme, subject to optimization criteria and constraints. In this effort we use modified greedy heuristics from the CRPP and investigated the proposed mechanism using a trace driven java based simulation.**

***Index Terms*—Replica placement problem, availability awareness, distributed objects, failure resilient.**

## I. INTRODUCTION

Replication is used in order to increase the availability, performance and fault tolerance of a distributed. Examples include video servers [1], [2], distributed databases [3], [4], distributed file systems [5], content distribution networks (CDNs) [6], Grid [7]and Cloud [8]. Critical data management issues for the success of the above systems in context of replication are how to create, update, access and distribute replicas of objects. The replica placement problem (RPP), in literature it is also known as file allocation problem [9]is intended to address the mention data management issues. In fact the first formulations for replica placement problem set back to early 70's [10], with the development of distributed computing systems spanning wide area networks, the interest on replica placement problem was reincarnated (see [11]–[13] for recent publications).

A generic replica formulations fall into two categories: static and dynamic. . Majority of the static replica placement variations believes that access profile do not change, and hence the replica placement scheme computed once remains for a large time span. In contrast to this reason, they do not incorporate the cost of replica creation, as it would be amortized by the large time span [1], [9], [10], [14]–[17]. Such a formulations is named as 1RPP by [18].

On the other hand dynamic formulations e.g., [3], [24], [28], change the replica placement scheme possibly upon every request. Indeed, they are more beneficial when the considered objects to replicate are relatively small in size and replication is done at on points along the request path [28]. However, if the objects are of larger sizes and availability is concern, dynamic schemes become less useful, e.g., distributed video servers [5]. Summarizing, the fact [18] classified the static solutions act as push-based prefetching schemes, and the dynamic ones as pull-based methods.

The assumption made so far in the literature, is whenever there is a need to recalculate the replica placement scheme, due to changes in user preferences, one of the 1RPP algorithms can be used to obtain a new solution. However, this approach does not consider the difficulties associated with the required object transfers, in transition from one replica placement scheme to another. To overcome this drawback Continuous Replica Placement Problem (CRPP) is proposed in [18] and further extended by that allows for more frequent updates on the replication scheme and in lights an underlying scheduling problem.

CRPP and RPP solutions are useful in order to guarantee minimum availability requirements; this does not entail they should remain unchanged. The collateral assumption made so far in the literature, is that the system is failure free. Here, we demonstrate that the existing approaches replica placement and its extensions does not consider the in depth and the difficulties associated presumably due to the node failure, link failure or any cause of node or component unavailability with performing the necessary replica management actions, in order to deploy or move from one replication scheme to another. Therefore, we propose an extension to CRPP formulations, called Availability Aware Continuous Replica Placement Problem (AACRPP) that allows for a more failure resilient replication scheme and the underlying scheduling problem.

Our contributions include the following: i) we formulate

*This work is fully funded and partially funded by Bright Spark Program and High Impact Research Grant from the University of Malaya under reference BSP/APP/1635/2013 and UM.C/625/1/HIR/MOE/FCSIT/03 respectively.*

AACRPP and identify the underlying details of the problem, ii) we demonstrate how to modify existing algorithms for CRPP in order to make them work in AACRPP iii) developed a simulation environment for evaluating replica placement algorithms and their network cost.

The rest of the article is organized as follows. Section 2 formulates AACRPP. Section 3 illustrates the heuristics, while Section 4 presents the performance evaluation. An overview of the related work is included in Section 5. Finally, Section 6 discusses some concluding remarks and demonstrates future work directions.

## II. Problem Formulation

In this section AACRPP is formalized by extending the CRPP. First the underlying assumptions on the system model are described followed by the problem formulation.

### A. System Model

Consider a generic distributed system where failure can occurs. The system is composed of $M$ servers also called as nodes and $N$ objects also called as items. Let $s(S_i)$ represent total storage capacity of a server $S_i$, where $1 \le i \le M$. Also, $s(O_k)$ denotes the storage size of object $O_k$, where $1 \le k \le N$.

The underlying network topology is a general graph, the communication across two servers $S_i$ and $S_j$ follows the shortest path formulation either via direct via point-to-point links (if any), or indirect via other servers. Intuitively, the per-byte cost over the network between the two servers $S_i$ and $S_j$ is denoted by $l_{ij}$, where we assume that $l_{ji} > 0$ for $j \ne i$, $l_{jj} = 0$ and $l_{ji} = l_{ij}$.

We assume that every single object $O_k$ has at least one stored replica known as primary replica hosted on a designated server, identified by $P_k$. Additional copies of $O_k$ may be hosted on other servers. If an object $O_k$ is hosted on a server $S_i$, then $S_i$ is known as *replicator* of $O_k$. The replica placement scheme of the system is coded in the form a matrix referred as the *replication matrix* denoted by letter $X$ with *MxN* dimensions. Where element $X_{ik}$ is 0 if a server $S_i$ is not a replicator of object $O_k$ and 1 otherwise. Intuitively, a replication matrix is *valid* if it holds the following two constraints:

$$\sum_{k=1}^{N} X_{ik} s(O_k) \le s(S_i), \forall i \quad \text{(Server storage constraint) (1)}$$

$$P_k = i \Rightarrow X_{ik} = 1, \forall k \quad \text{(Primary replica constraint) (2)}$$

The client object request processing flow is modeled in Fig. 1. In this model any server $S_i$ might receive a user/ client request for accessing any system object $O_k$. If server $S_i$ is a replicator of the requested object $O_k$ then the request is furnished directly. Otherwise, $S_i$ redirect the request to the nearest server in terms of network cost $S_j$ derived through a function $N_{ik}^{X}$, where $X_{jk} = 1$. In this case the cost for the server network is proportional to the data (size of request plus size of reply) exchanged between $S_i$ and $N_{ik}^{X}$, the respective link cost $l_{iN_{ik}^{X}}$. For the purpose of symmetry, if a server $S_i$ is a replicator of object $O_k$, $N_{ik}^{X} = i$. The notation $N_{ik}$ is used in favor of readability instead of $N_{ik}^{X}$.

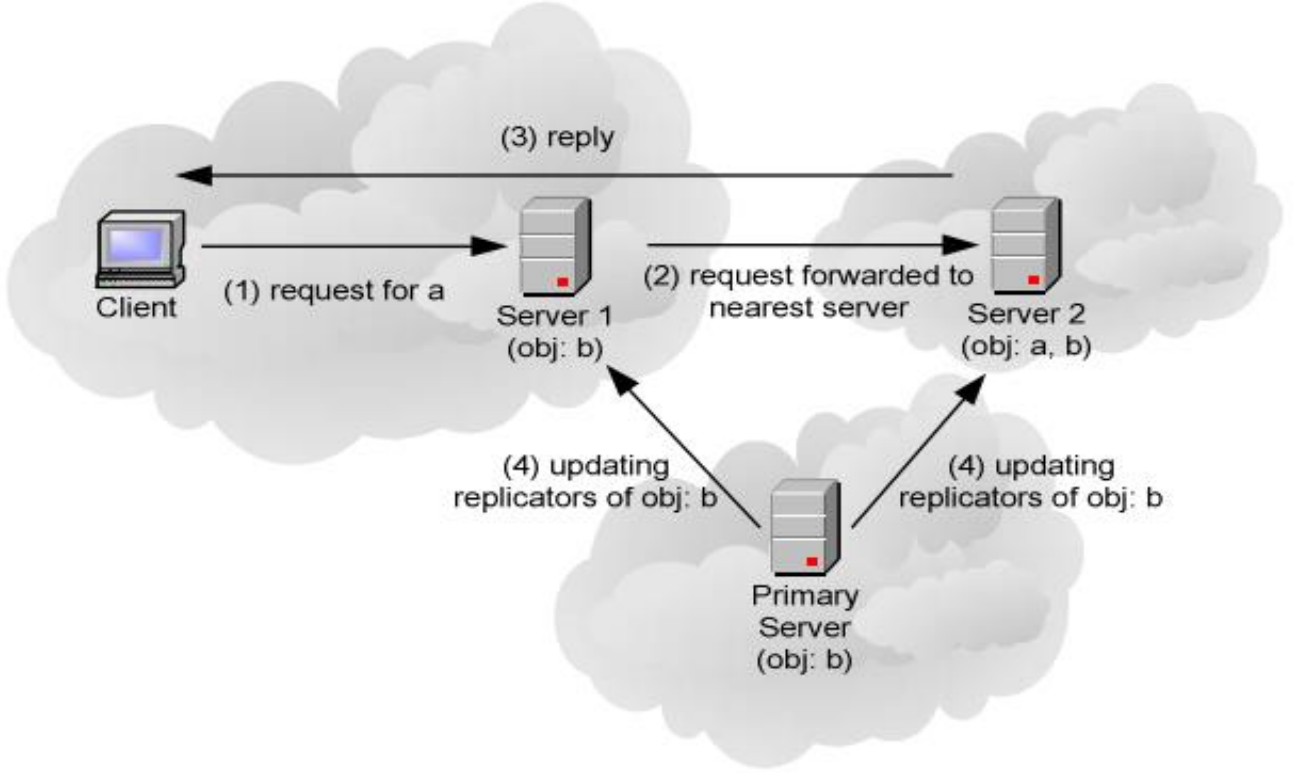

Fig. 1 System Model and Client Request Processing Flow (source [18] )

Let $r_{ik}$ denote the total client traffic flow (request and response) in bytes for object $O_k$ at server $S_i$. The network cost incurred to furnish the requests for object $O_k$ at server $S_i$ is calculated as $l_{iN_{ik}^{X}} r_{ik}$ (since $X_{ik} = 1 \Rightarrow l_{iN_{ik}^{X}} = 0$). Respectively, the cost of all requests for object $O_k$ at all servers in the system is:

$$R_k^X = \sum_{i=1}^{M} l_{iN_{ik}^{X}} r_{ik} \quad (3)$$

And the total network cost due for all requests of all objects at all servers in the system is:

$$C^X = \sum_{k=1}^{N} R_k^X \quad \text{or} \quad C^X = \sum_{k=1}^{N} \sum_{i=1}^{M} l_{iN_{ik}^{X}} r_{ik} \quad (4)$$

For simplicity, we will refer $R_k^X$ and $C^X$ as $R_k$ and $C$, respectively.

Let $Pf_i$ denote the availability probability of a server $S_i$. Consequently, the availability probability of an object $O_k$ is defined as:

$$Pf_k^{'} = \prod Pf_i \, , \; \forall i, \; \text{where} \; X_{ik} = 1 \quad (5)$$

### B. Availability Aware Continuous Replica Placement Problem (AACRPP)

Consider a generic distributed system implementing a replica placement scheme $\boldsymbol{X^{old}}$. Changes in user requests patterns and server availability profile may result in the need to define a new replica placement scheme $\boldsymbol{X^{new}}$. Where $C^{X^{old}}$ and $C^{X^{new}}$ be the access costs (derived from (4)) for $\boldsymbol{X^{old}}$ and $\boldsymbol{X^{new}}$ respectively. We define a maximization function by extending the benefit function of [18] in order to decide whether the new replica placement scheme should be implemented:

$$B^{X^{old}, X^{new}} = (C^{X^{old}} - C^{X^{new}} - I^{X^{old}, X^{new}}) \times Pf_i \quad (6)$$

$C^{X^{old}} - C^{X^{new}}$ is the difference of cost between the existing replication state and the new one, $I^{X^{old},X^{new}}$ is the respective implementation cost and $Pf_i$ is the availability probability of a server $S_i$ which is currently the focus to add an object to its storage.

In order to maintain the availability of an object $k$ which is the focus to be added to a server the following condition should also met:

$$Pf_k^{\prime new} \geq Pf_k^{\prime old} \quad (7)$$

$Pf_k^{\prime \boldsymbol{new}}$, *and* $Pf_k^{old}$ is the availibility probability of object $k$ for $\boldsymbol{X^{old}}$ and $\boldsymbol{X^{new}}$ respectively.

*For a given client request traffic profile, a server failure profile and a current replication sate matrix, find a new replication state matrix such that the benefit function (6) is maximized, while meeting the storage constraint (1), primary replica constraint (2), and object availability constraint (7).*

## III. Heuristics For AACRPP

In this section an algorithms based on the greedy paradigm is presented, i.e., GreedyGlobal (GG). Variants of the algorithms were proposed earlier in the literature [18] in order to tackle RPP and CRPP.

### A. *Availability Aware Greedy Global (AAGG)*

This algorithm starts with an initial replication state matrix $X^{old}$ and follows iteratively to produce a new replication state $X^{new}$. The pseudocode of this greedy techniques is given below:

**Algorithm 1 Availability Aware Greedy Global**

```
X: = Xold; b: = 0;
While POSITIVE_FLIP_EXISTS () do
    Repeat
    X':= X;
        (i,k) := FIND_UNMARKED_POSITIVE_FLIP();
        MARK_TEMP_FLIP (i, k);
        While !HAS_SPACE_FOR(i,k) do
            k' := -1; b' := MIN_INTEGER;
            For k'' := 1 to N
                If X' [i][k''] = 1 &&Pk'' != i then
                    X'[i][k''] := 0; cost := 0;
                    b'' := AACRPP_BENEFIT_FUNCTION(X,X',cost);
                    If b'' > b' then k':= k''; b':= b''; End
                    X'[i][k''] := 1;
                End
            End
            If k' = -1 then break; End // cannot free more space
            X'[i][k'] :=0;
        End
        If HAS_SPACE_FOR(i,k) and Pf'X'k ≥ PfXk then
            j := Nik;
            X'[i][k] := 1; cost := s(Ok)lij;
            b' := AACRPP_Benefit_Function(X,X',cost);
            If b'> b then bi := i; bk := k; b := b'; Xb := X'; End
        End
    Until !UNMARKED_POSITIVE_FLIP_EXISTS();
    If b = 0 then break; End // all options lead to a worse situation
    X:= Xb; b:= 0;
    UNMARK_TEMP_FLIPS ();
End
return X;
```

In every iteration, all possible positive element flips $X_{ik} \rightarrow 1$, are considered, and the one that respects the constraints (1), (2) and (7) while maximizing the benefit function (6) is chosen. Since the only change that occurs is server $S_i$ became a replicator object $O_k$ the implementation cost of this new placement is equal to the cost for requesting the object from the nearest replicator: $l_{iN_{ik}} s(O_k)$. If the storage capability of server $S_i$ is not sufficient to store object $O_k$, other object replicas $O_{k'}$ hosted on $S_i$ are deleted in increasing order of the benefit function (6). The heuristics converges when there is no sufficient storage capacity in each server to add a new replica or any further replica creation leads to a negative benefit.

### A. *Availability Aware Greedy Random Object (AAGRO)*

The Availability Aware Greedy Random Object algorithm is similar to Availability Aware Greedy Global, but focuses on the replication of the same object at a time. The pseudo code follows:

**Algorithm 2 Availability Aware Greedy Random Object**

```
X: = Xold;
While NOT_ALL_OBJECTS_CONSIDERED() do
    k := RANDOM_PICK_OBJECT();
    While POSITIVE_FLIP_EXISTS_FOR(k) do
    // same as AAGG, but with fixed k
    End
End
return X;
```

The algorithm initiates by randomly picking an object $O_k$ and proceed in the same fashion as AAGG for this particular object. More precisely, a single replica allocation is performed in each iteration, until no more beneficial replicas can be allocated. The next object is chosen at random, and the same procedure is repeated. The algorithm converges when every object has been considered.

## IV. Evaluation

In order to evaluate the feasibility of our problem we developed a generic java based cross platform trace driven simulation environment for evaluating replica placement algorithms and its variants called TestRPA. The abstract block diagram of the simulation environment is presented in Fig. 2:

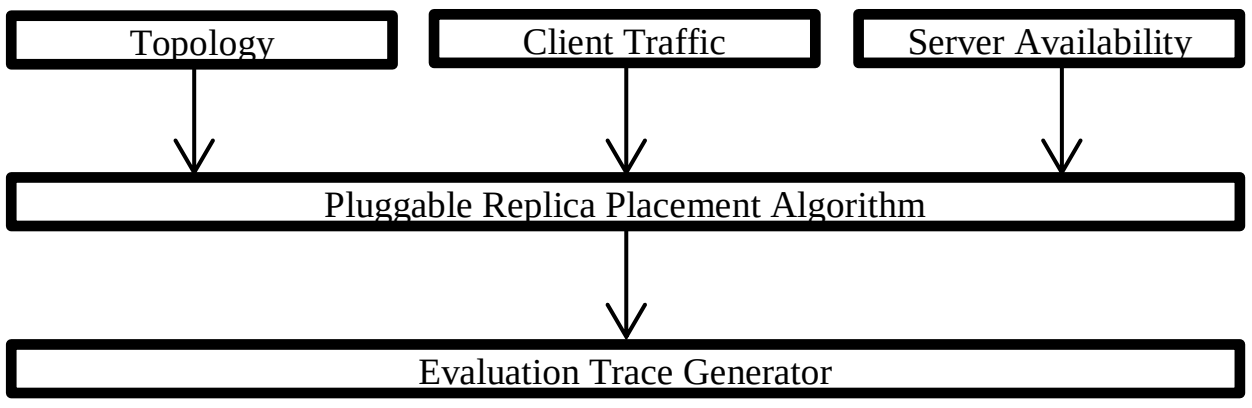

Fig. 2 Abstract Block Diagram of Simulation Environment

The environment can either evaluate replica placement algorithm for distributed object based systems like CDNs and Distributed Databases or can derive network cost also called the cost of the schedule between two consecutive replication states. Currently we are defining scenarios and metrics other than the implementation cost to evaluate the proposed algorithms as the availability metrics included in our formulation changed the essence of the problem. A general principle of computing and multi attribute optimization problem is when one attribute is optimized the other ones usually has a negative impact and sometimes a positive also.

### A. Setup

The server topology was generated using BRITE [19], for 50 nodes each having a connectivity of 1. Node connections are following the Barabasi-Albert model, which has been used to describe power-law router graphs [20]. Links were assigned a fixed cost, uniformly distributed between 1 and 10. Point-to-point communication costs were set equal to aggregated link cost along the shortest (less costly) paths. A set of 1000 objects was used, with sizes uniformly distributed between 1000 and 5000. The primary replicas were randomly assigned to the server nodes. The server availability probability is calculated from the failure trace provided in [21].

As we are in preliminary stage of our work does not identified metrics for comparison other than the implementation cost of the schedules produced by the CRPP Heuristics and AACRPP. Fig. 3 plot the cost of the schedules produced by CRPP and AACRPP heuristics for our first experimental setup. The new replica placement scheme produced by these heuristics is constrained by the number of replicas per object and is controlled during the course of experiments. The storage capacity of each server is set equal to the sum of the replicas it must host in $X^{new}$.

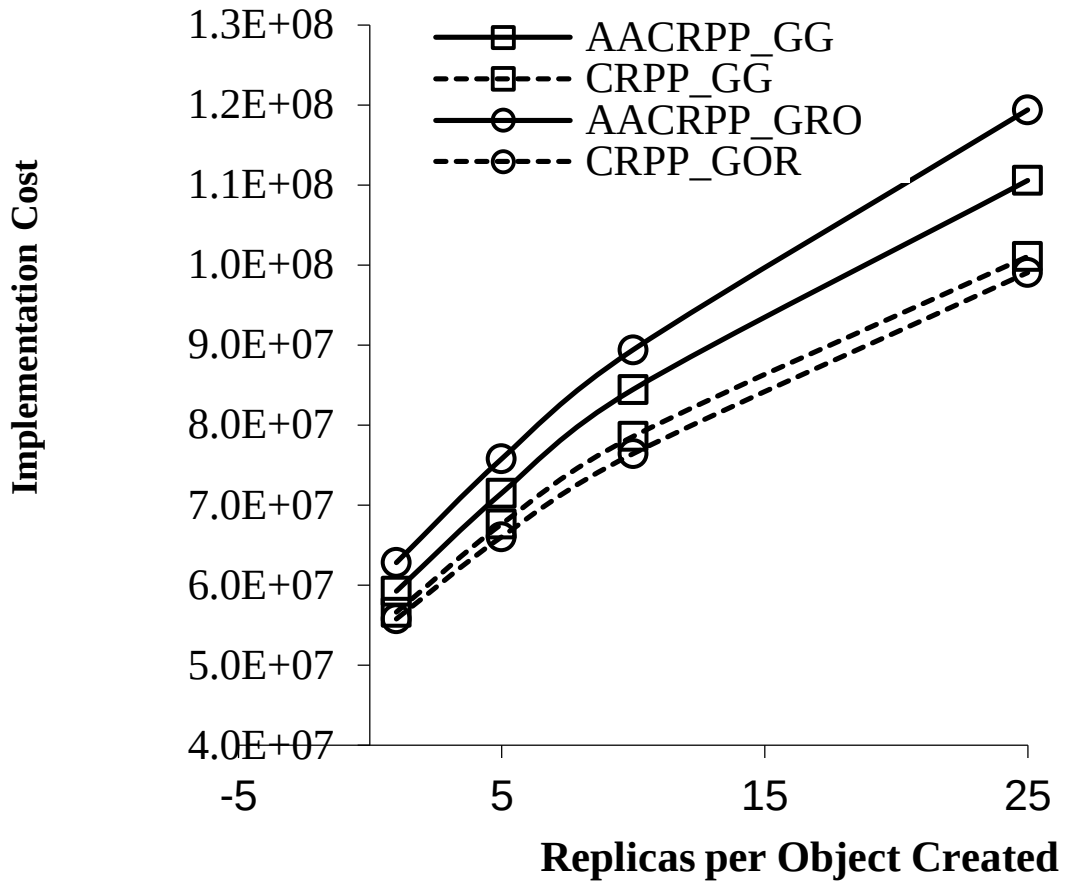

Fig. 3 Schedule cost of AACRPP and CRPP Based heuristics, while increasing number of replicas per object starting from primary copies only

## V. Related Work

The Replica Placement Problem has been extensively researched, and a plethora of problem definitions are available in this context. In [22] client-replica distance is considered as the optimization target, the primary goal of [23] is load balancing. Access cost is the focus in [24]–[26]. Other issues taken into account in conjunction with RPP formulations are server storage capacity [27], [28], processing capacity [29] and bandwidth [30] to name a few. In this article we have adopted a model similar to [18]. The first insight to minimize the implementation cost and derived a replica scheme from the existing replica placement is provided in [18]. Although our AACRPP definition is an extension to [18] however can be extended to include additional parameters, in this work we have focused in what we believe to be the essence of the problem. To the best of my knowledge I am unaware of any previous research efforts that try to place replicas on the basis of availability awareness of nodes and objects.

## VI. Conclusions And Future Work

In this paper we formulated the case of availability aware continuous replica placement. In this effort, we tailored the CRPP heuristics to handle the formulated case and provided preliminary results by a custom simulation environment. However CRPP includes an underlying scheduling task which is not underlined in this work and does not formulated the scheduling in context of availability awareness.

Future work includes an extended version where we evaluate our approach by varying the network topologies, traffic traces, server storage capacities and failure models against a distributed system scenarios such as distributed cloud storage networks and extending the concept of availability awareness replica placement to the data center topologies and rack aware availability conscious data placement. Further the CRPP scheduling task is to be formulated to incorporate an availability aware schedule of object transfers. Moreover, the topological changes and object changes is also an area of interest in the future.

## Acknowledgment

This work is fully funded and partially funded by Bright Spark Program and High Impact Research Grant from the University of Malaya under reference BSP/APP/1635/2013 and UM.C/625/1/HIR/MOE/FCSIT/03 respectively.